\documentclass{nature}
\usepackage{times}
\usepackage{url}
\usepackage{xcolor}
\usepackage{hyperref}
\usepackage{amsmath}
\usepackage{amssymb}
\usepackage{amsfonts}
\usepackage{graphicx}
\graphicspath{{figs/}}
\usepackage{multirow}
\usepackage{dcolumn}
\newcolumntype{d}[1]{D{.}{.}{#1}}
\makeatletter
\let\saved@includegraphics\includegraphics
\AtBeginDocument{\let\includegraphics\saved@includegraphics}
\makeatother

\newcommand{\si}[1]{\textcolor{blue}{\textit{SI} #1}}

\title{\large Dynamics of Cross-Platform Attention to Retracted Papers}
\author{Hao Peng$^{1}$, Daniel M. Romero$^{1,2,3}$, Em\H oke-\'Agnes Horv\'at$^{4,5,6}$}

\begin{document}

\baselineskip21pt
\maketitle 

\begin{affiliations} 
\item School of Information, University of Michigan, Ann Arbor, MI 48109, USA
\item Center for the Study of Complex Systems, University of Michigan, Ann Arbor, MI 48109, USA
\item Department of Electrical Engineering and Computer Science, University of Michigan, MI, USA
\item School of Communication, Northwestern University, Evanston, IL 60208, USA
\item McCormick School of Engineering, Northwestern University, Evanston, IL 60208, USA
\item Northwestern University Institute on Complex Systems, Evanston, IL 60208, USA
\end{affiliations} 

\begin{abstract}
Retracted papers often circulate widely on social media, digital news and other websites before their official retraction. The spread of potentially inaccurate or misleading results from retracted papers can harm the scientific community and the public. 
Here we quantify the amount and type of attention 3,851 retracted papers received over time in different online platforms. Comparing to a set of non-retracted control papers from the same journals, with similar publication year, number of co-authors and author impact, we show that retracted papers receive more attention after publication not only on social media, but also on heavily curated platforms, such as news outlets and knowledge repositories, amplifying the negative impact on the public. 
At the same time, we find that posts on Twitter tend to express more criticism about retracted than about control papers, suggesting that criticism-expressing tweets could contain factual information about problematic papers.
Most importantly, around the time they are retracted, papers generate discussions that are primarily about the retraction incident rather than about research findings, showing that by this point papers have exhausted attention to their results and highlighting the limited effect of retractions. 
Our findings reveal the extent to which retracted papers are discussed on different online platforms and identify at scale audience criticism towards them. In this context, we show that retraction is not an effective tool to reduce online attention to problematic papers.
\end{abstract}

\section*{Introduction}
Retraction in academic publishing is an important and necessary mechanism for science to self-correct \cite{ajiferuke2020correction}.
Prior studies show that the number of retractions has increased in recent years \cite{van2011science,steen2011retractions,fang2012misconduct,shuai2017multidimensional}.
This rise can be explained by many different factors  \cite{budd1998phenomena,steen2011retractions,fang2012misconduct,steen2013has,shuai2017multidimensional}.
One reason is that the number of publications is increasing exponentially \cite{dong2017century}.
Meanwhile, as scientific research has become more complex and interdisciplinary than ever before, reviewers are facing a higher cognitive burden \cite{lee2013bias,jones2008multi}.
This undermines the scientific community's ability to filter out problematic papers.
In fact, research shows that prominent journals with rigorous screening and high publishing standards are as likely to publish erroneous papers as less prominent journals \cite{cokol2007many}.
Finally, not all retractions are due to research fraud---some papers are retracted due to unintentional errors or mistakes, which become more likely as research data grows in size and complexity \cite{budd1998phenomena}.

Regardless of the reasons behind this increase,
a high incidence of retractions in academic literature has the potential to undermine the credibility of scientific communities and reduce public trust in science \cite{budd1998phenomena,van2011trouble}. 
What's more, the circulation of misleading findings can be harmful to the lay public \cite{budd1998phenomena,marcus2018scientist,sharma2020patterns,grady2021honesty}, especially given how broadly papers can be disseminated via social media \cite{ke2017systematic}.
For instance, there are two retracted papers among the 10 most highly shared papers in 2020 according to Altmetric, a service that tracks the online dissemination of scientific articles \cite{alt10top}.
One of them, published in a top biology journal, reported that treatment with chloroquine had no benefit in COVID-19 patients based on data that were likely fabricated \cite{mehra2020retracted}.
Another paper, published in a well-regarded general interest journal, falsely claimed that having more female mentors was negatively correlated with post-mentorship impact of junior scholars \cite{alshebli2020retracted}.
Both papers attracted considerable attention before they were retracted, raising questions about their possible negative impact on online audiences' trust in science.

As these examples suggest, retracted papers can attain substantial online attention and potentially flawed knowledge can reach the public who often are impacted by the research results \cite{clauset2017data}. 
This large-scale spreading of papers occurs as the Web has become the primary channel through which the lay public interacts with scientific information \cite{sundar1998effect,flanagin2000perceptions,scheufele2019science}. 
Past research on the online diffusion of science has mainly studied the spread of papers without regard to their retraction status \cite{morris2012tweeting,brossard2013new,scheufele2013communicating,milkman2014science,su2015science,zakhlebin2020diffusion}.
Other work has examined the dissemination of retracted papers in scientific communities, focusing mainly on the associated citation penalty \cite{lu2013retraction,madlock2015lack,shuai2017multidimensional,sharma2020patterns,avenell2019investigation,budd2016investigation,wadhwa2021temporal,jin2019reverse,azoulay2015retractions}. 


\begin{figure*}[ht!] 
\centering
\includegraphics[trim=0mm 0mm 0mm 0mm, width=0.9\linewidth]{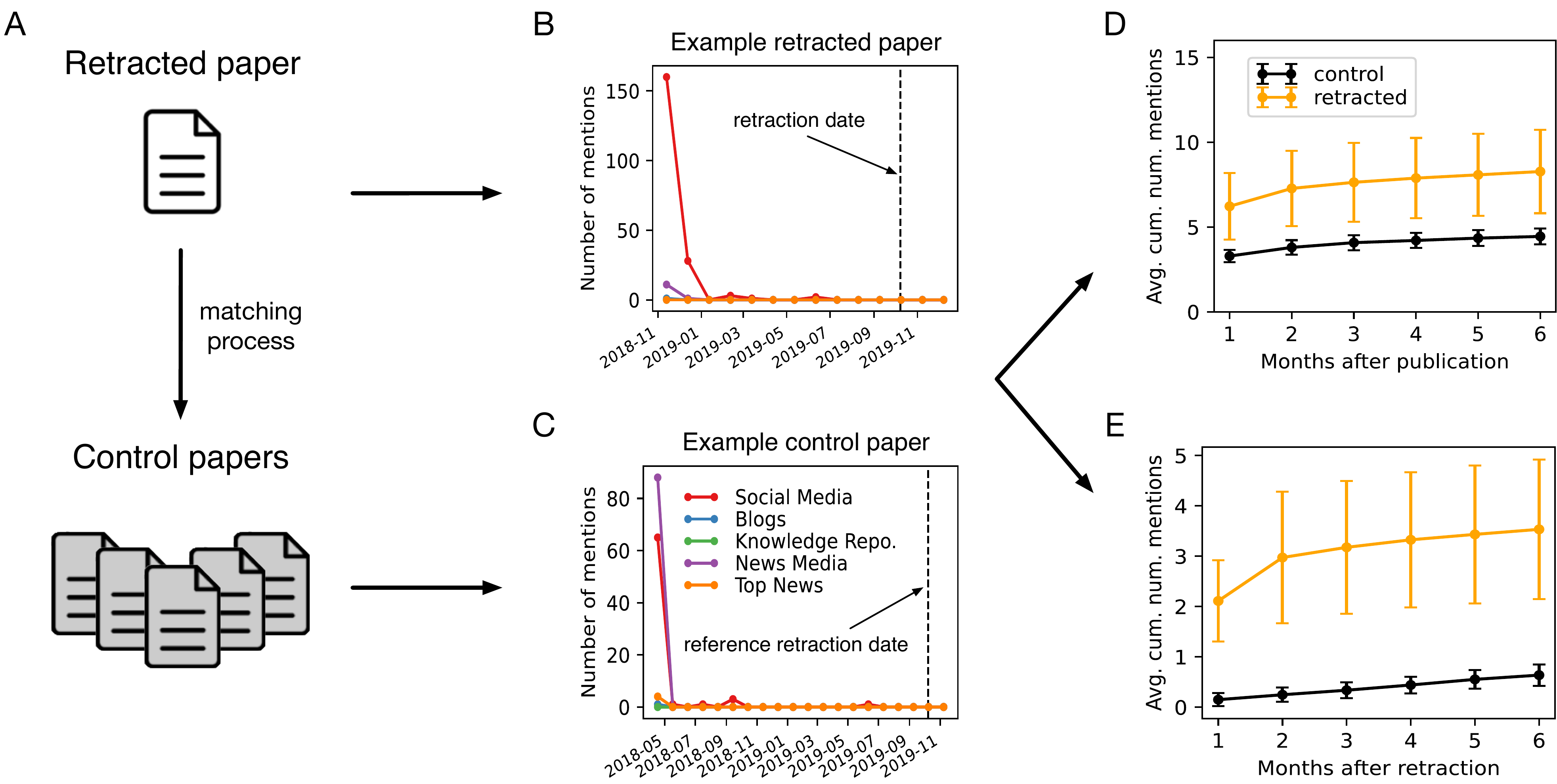}
\caption{\textbf{Illustration of the research process that compares the online attention received by retracted and control papers.} \textbf{A}, We match five control papers to each retracted paper using the Altmetric database. \textbf{B-C}, We track the change in mentions over time on four different types of platforms and in top news outlets. As an example, we show here the retracted paper ``Effect of a Program Combining Transitional Care and Long-term Self-management Support on Outcomes of Hospitalized Patients With Chronic Obstructive Pulmonary Disease: A Randomized Clinical Trial'' published in \textit{JAMA} (DOI: 10.1001/jama.2018.17933) and one of its matched control papers ``Vitamin D, Calcium, or Combined Supplementation for the Primary Prevention of Fractures in Community-Dwelling Adults: US Preventive Services Task Force Recommendation Statement'' (DOI: 10.1001/jama.2018.3185). \textbf{D}, We compute the average cumulative number of mentions across all platforms within 6 months after publication (and before retraction) for all retracted and control papers in the dataset. \textbf{E}, Similarly, we compute the average cumulative number of mentions within 6 months after retraction. Error bars indicate 95\% confidence intervals.}
\label{fig-one}
\end{figure*} 

However, the impact of retraction on the online dissemination of retracted papers is unclear. Here, we address this essential open question. Past research found that authors tend to keep citing retracted papers long after they have been red-flagged, although at a lower rate \cite{budd1998phenomena,van2011trouble,shuai2017multidimensional,sharma2020patterns}. This raises the question of whether retraction is effective in reducing public attention beyond the academic literature. Studying the impact of retraction relative to the temporal ``trajectory'' of mentions a paper receives could be helpful for journals to devise policies and practices that maximize the effect of retractions \cite{van2011trouble,brainard2018massive}.

To understand whether retraction is appropriate for reducing dissemination online, we first assess the extent of the online circulation of erroneous findings, by investigating variations in how often retracted papers are mentioned on different types of platforms before and after retraction. Recent research indicates that, overall, retracted papers tend to receive more attention than non-retracted ones \cite{serghiou2021media}. Prior work also shows that retractions occur most frequently among highly-cited articles published in high-impact journals \cite{cokol2007many,furman2012governing}, suggesting a counter-intuitive link between rigorous screening and retraction. Is there a similar tendency online where retracted papers receive more attention on carefully curated platforms such as news outlets than on platforms with limited entry barriers like social media sites? Such a trend would highlight difficulties with identifying unreliable research given their broad visibility in established venues, and could inform attempts to manage the harm caused by retractions. 
Second, we distinguish between critical and uncritical attention to papers to uncover how retracted research is mentioned. More than half of retracted papers are flagged because of scientific misconduct such as fabrication, falsification, and plagiarism \cite{fang2012misconduct,shuai2017multidimensional}. These papers may receive lots of attention due to criticism raised by online audiences. 
Is the attention received by retracted papers due to sharing without knowing about the mistakes of a paper or is it rather expressing concerns (so-called ``critical'' mentions)? 
As suggested in a recent case study \cite{haunschild2021can}, knowing how retracted papers are mentioned may uncover users who are improving science-related discussions on Twitter by identifying papers that require a closer examination. 


In this paper, we compiled a dataset to quantify the volume of attention that 3,985 retracted papers received on 14 online platforms, e.g., public social media posts on Twitter, Facebook, and Reddit, their coverage in online news, citations in Wikipedia, and research blogs. We compared their attention with non-retracted papers selected through a matching process based on publication venue and year, number of authors, and author's citation count (Figure~\ref{fig-one}A and ``Identifying control papers'' in \nameref{methods}). We obtained retracted papers from Retraction Watch \cite{rw}, the largest database to date that records retracted papers, and pulled their complete trajectory of mentions over time on various platforms from a service called Altmetric \cite{alt} that has been tracking posts about research papers for the past decade.
The granularity and scale of the data enabled us to differentiate mentions on four different types of platforms, including social media, news media, blogs, and knowledge repositories (Figures~\ref{fig-one}B-C). 
We thus provide the first systematic investigation of the online mentions of papers disentangled by platform during the time periods between publication and retraction (Figure~\ref{fig-one}D), and after retraction (Figure~\ref{fig-one}E). 

Our findings offer insights into how extensively retracted papers are mentioned in different online platforms over time and how frequent their critical vs uncritical discussion is on Twitter.
Most importantly, we show that retractions are not reducing harmful dissemination of problematic research on any of the platforms studied here, because by the time the retraction is issued, most papers have exhausted their online attention. 
We also contribute a large dataset of identifiers of tweets that mention the papers used in this study and human annotations of whether the tweets express criticism with respect to the findings of the papers (``Labeling critical tweets'' in \nameref{methods}). This dataset can be useful to the broader research community for the study of criticism towards scientific articles, and can aid the development of automated methods to detect criticism computationally \cite{pei2021measuring}.

\section*{Are Retracted Papers More Popular Even in the News?}

To understand whether retraction can contain the spread of problematic papers online in comparison with control papers, we first evaluate the amount of attention that retracted papers receive over time on four types of platforms. Then, we identify at large scale discussions on Twitter that criticize papers before they are retracted. 

\begin{figure*}[ht!] 
\centering
\includegraphics[trim=0mm 0mm 0mm 0mm, width=0.85\linewidth]{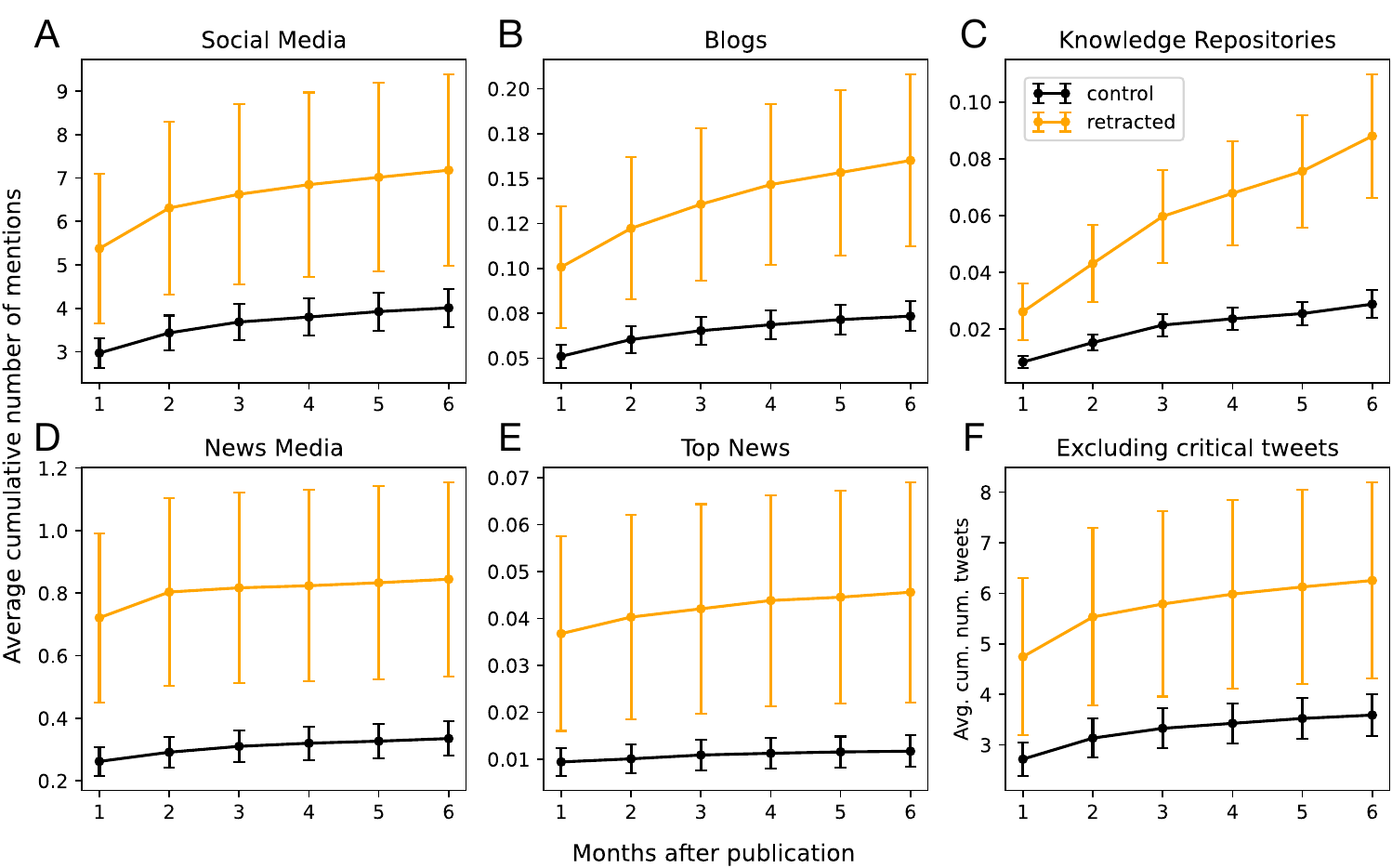}
\caption{\textbf{After publication and before retraction mentions.} \textbf{A-E}, Average cumulative number of mentions received within 6 months after publication on four types of platforms and in top news outlets for both retracted and control papers. \textbf{F}, Average cumulative number of mentions on Twitter after excluding critical tweets. Comparisons across different types of platforms show that retracted papers receive more attention after publication than non-retracted papers. Error bars indicate 95\% confidence intervals.}
\label{fig-after-pub}
\end{figure*} 

Retracted papers attract more overall attention than control papers (Figure~\ref{fig-one}D). However, does this trend apply to all types of platforms?
As shown in Figures~\ref{fig-after-pub}A-E, across 2,830 retracted and 13,599 control papers with a tracking window of at least 6 months (see ``Defining tracking windows'' in \nameref{methods}), we find that retracted papers receive more attention after publication on all four types of platforms and are also mentioned more in news outlets with high-quality science reporting (see \si{Methods} for outlet selection).
On average, papers obtain mentions most frequently on social media, followed by news media, and they receive roughly similar amounts of attention on blogs and knowledge repositories. Changing the length of the tracking window produces qualitatively similar results (\si{Fig. S1}). 

The distribution of online attention to retracted papers is right-skewed, meaning that most papers do not receive much attention while a few become very popular (\si{Fig. S2}). 
To statistically compare the attention between retracted and control papers while controlling for fundamental factors that could affect the amount of attention a paper receives, we performed negative binomial regression to examine the association between retraction and attention.
The association between a higher chance of retraction and more mentions of a paper is significant across all types of platforms and across different time windows (\si{Table S1}).
Furthermore, a Mann–Whitney U test also shows that the central tendency of the distribution of mentions for retracted papers is larger than that of control papers on all types of platforms but news media (\si{Table S2}). Since matching based on journals in identifying control papers is imperfect in accounting for research topics, particularly for papers published in multidisciplinary journals, we repeated the analysis by excluding papers in general science journals and found our results to be robust to this change (see \si{Fig. S13}). 

Moreover, investigating the ratio of the average number of mentions of retracted papers relative to their control counterparts, we find more attention to retracted papers in news outlets, including top news, and knowledge repositories than in social media and blogs (\si{Fig. S3}). This suggests that curated content tends to contain more mentions of retracted findings for each mention of non-retracted research than unfiltered contributions by individuals in social media.

\section*{Are Retracted Papers Shared Critically?}

Since it is possible that the additional mentions of retracted papers are not meant to propagate their findings but to express criticism, we repeated the analysis after excluding posts that express concerns about the claims of the paper.
Here, we focused on Twitter, the single largest platform in our dataset, which accounts for about 80\% of all posts in the Altmetric database. Twitter features a diverse representation of different types of users who share scientific content, including academics, practitioners, news organizations, and the lay public \cite{kwak2010twitter,ke2017systematic} (\si{Fig. S16}).

We define a \textit{critical tweet} as a tweet expressing uncertainty, skepticism, doubt, criticism, concern, confusion, or disbelief with respect to a paper's findings, data, credibility, novelty, contribution, or other scientific elements.
We adopt the term ``uncritical attention'' to refer to mentions of the paper that do not express criticism.
We used a combination of automated and manual methods to label critical tweets with high recall. We identified a large number of critical tweets after collecting 3-4 independent expert annotations for thousands of tweets (see ``Labeling critical tweets'' in \nameref{methods}).

Figure~\ref{fig-after-pub}F shows that even after excluding critical tweets for both groups of papers (see full regression results in \si{Table S3}), retracted papers still receive more mentions on Twitter than control papers, suggesting that potentially flawed findings are indeed uncritically mentioned and shared more. 
The number of tweets does not necessarily reflect a paper's true influence since different posts may have different audience sizes. Based on the number of followers of users who mention retracted vs control papers, the former indeed reach comparable or even more people after publication than the latter (\si{Fig. S4}).
Similar findings may also apply to other platforms such as news outlets, given that journalists are discouraged from publishing news articles that question a scientific paper \cite{rodder2011sciences, blum2006field}.



\begin{figure*}[ht!] 
\centering
\includegraphics[trim=0mm 0mm 0mm 0mm, width=0.5\linewidth]{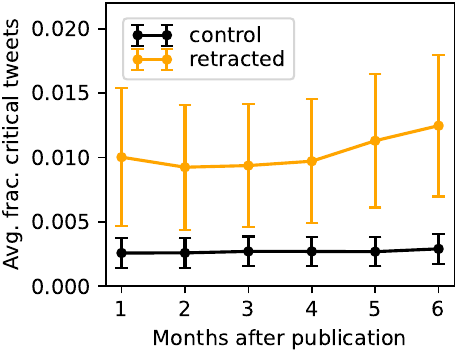}
\caption{\textbf{Average cumulative fraction of critical tweets within 6 months after publication.} Control papers are selected from a matching process that considers the publication year, publication venue, number of authors, and author's citation count. 
We focused on papers that have at least one tweet mention in each time window.
Retracted papers receive a higher fraction of critical tweets.}
\label{fig-after-pub-uncrtn}
\end{figure*}

Having examined uncritical attention to papers on Twitter, we now analyze their critical attention.
A recent case study suggested that Twitter attention received after publication could be indicative of factual information that should be investigated \cite{haunschild2021can}.
To assess this possibility at a large scale, we investigated how frequently retracted papers receive critical tweets between publication and retraction. For control papers, we identified critical tweets between publication and the retraction date of their matched paper (see an illustration of the reference date in Figure~\ref{fig-one}C).
Figure~\ref{fig-after-pub-uncrtn} shows that, compared to control papers, retracted papers receive a higher fraction of critical tweets, especially when looking at longer time periods, between 5 and 6 months. 
This finding is also supported by regression analyses with a paper's fraction of critical tweets as the dependent variable and the retraction status as the independent variable, while controlling for the publication year, the number of authors, and the author's log citations. The coefficients of the retraction status are positive and statistically significant ($p <$ 0.001 for 5 to 6 months, \si{Table S4}).
Note that we did not test whether this finding holds on other platforms aside of Twitter. Establishing a similar presence of critical discussions on other types of platforms requires further examination.


If papers that receive a high fraction of critical tweets are also largely ignored by news media and other platforms, then the critical signal from Twitter would not be as relevant. We measured the correlation between the fraction of critical tweets and the number of mentions on other types of platforms (\si{Table S5}), finding that there is no significant association. Retracted papers with a higher fraction of critical tweets are thus not necessarily those with more or less coverage in other types of platforms, such as news media.

Overall, this analysis suggests that Twitter readily hosts critical discussion of problematic papers well before they get retracted. These discussions credit voices that are actively helping to improve science-related discussions in digital media.

\section*{Is Retraction Effective in Reducing Attention?}

The fact that retracted papers are mentioned more often after their publication prompts us to ask if they continue to receive out-sized attention up until and even beyond their retraction. 
This leads to our main question focused on understanding the effectiveness of the retraction in reducing online attention.
We found that retracted papers are discussed much more frequently than control papers even over the 6 months after their retraction (\si{Fig. S5}). 
This is not unexpected, as it is very likely that people are reacting to the news of retraction in ways that involve mentioning the paper, such as questioning and criticizing the authors' practices, the peer review system, or the publishers. 

\begin{figure*}[ht!] 
\centering
\includegraphics[trim=0mm 0mm 0mm 0mm, width=\linewidth]{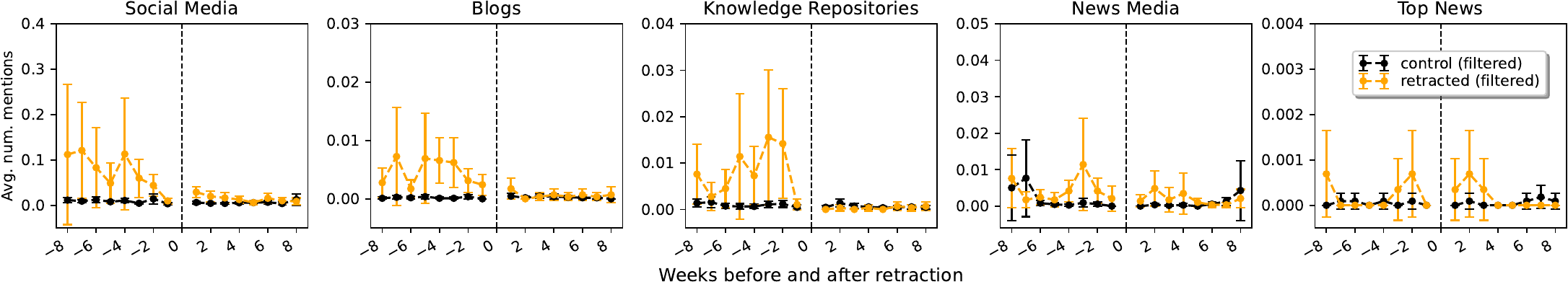}
\caption{\textbf{Average number of weekly mentions within 2 months before and after the retraction.} Trends are shown on four types of platforms and in top news outlets for data that exclude posts containing the phrase ``retract'' during the whole time period, for both control and retracted papers. 
In the case of retracted papers, we also manually excluded all after-retraction posts that discussed the retraction in some form. 
Comparisons across different types of platforms show that retracted papers do not receive statistically more mentions than control papers immediately before or after retraction.
Error bars indicate 95\% confidence intervals.}
\label{fig-before-after-ret}
\end{figure*}

To more accurately measure the extent to which potentially flawed results from retracted papers are shared without regard to their problematic nature, we excluded posts that discussed the retraction itself. To identify such content, we used a filtering strategy (see ``Filtering retraction-related posts'' in \nameref{methods}).
As \si{Fig. S5} shows, after this filtering, retracted papers are mentioned as much as or less than control papers. 
This result indicates that the additional attention to retracted papers after their retraction is primarily related to the retraction incident rather than to the paper's findings and the previously observed surplus of mentions to retracted papers has disappeared.
Yet, this finding does not show the effectiveness of the retraction itself. Since online attention tends to decay naturally over time \cite{wu2007novelty}, it is possible that retracted papers already received attention at a similar level as control papers by the time they were retracted. 
We thus compared the online attention to retracted and control papers before and after the retraction/reference date. 
Figure~\ref{fig-before-after-ret} shows that, across different types of platforms, retracted papers are not discussed significantly more often than control papers immediately before their retraction.
In fact, even without excluding any posts, 80.2\% of retracted papers receive no mentions over the 2 months preceding their retraction, while 93.6\% of them receive no mentions in the last month before retraction. This suggests that by the time the retraction is issued, most papers have already exhausted their attention, meaning that the retraction does not serve the purpose of further reducing the attention.
Note that we did not exclude critical tweets in this analysis, but only posts that discussed the retraction itself. If we had excluded critical posts before retraction for all platforms, this result would only become more stark since the pre-retraction mentions would decrease further, indicating even less attention that the retraction could intervene on.

\begin{figure*}[ht!] 
\centering
\includegraphics[trim=0mm 0mm 0mm 0mm, width=\linewidth]{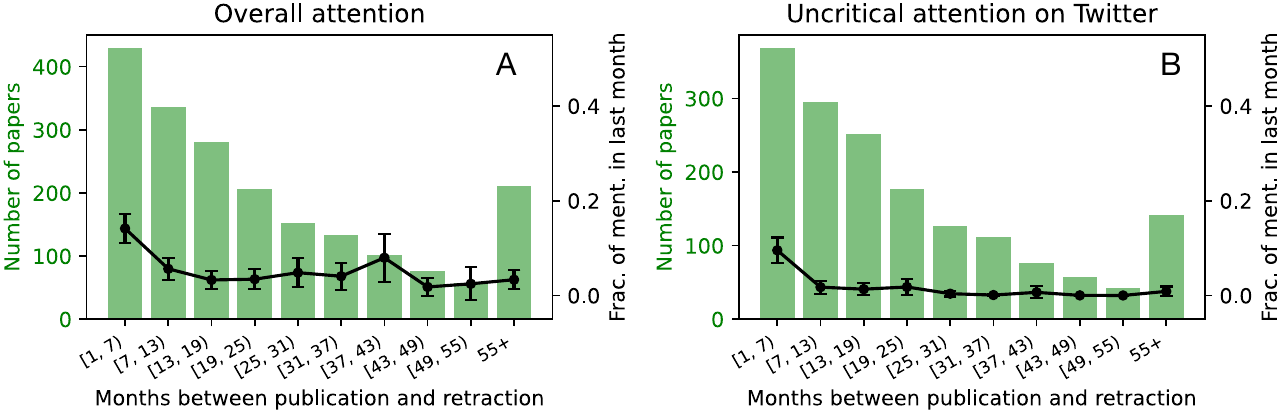}
\caption{\textbf{Average fraction of mentions in the last month before retraction.}
The x-axis represents different time windows between publication and retraction. 
The green bar plot (left y-axis) shows the number of papers in each time window, excluding those with 0 total mentions between publication and retraction.
The black line plot (right y-axis) shows the average fraction of mentions in the last month before retraction.
Error bars indicate 95\% confidence intervals.
\textbf{A}, Data based on mentions in all platforms. \textbf{B}, Similar results for uncritical mentions on Twitter.
}
\label{fig-last-month-frac}
\end{figure*}

We provide six robustness tests to further probe the finding that retraction has limited efficacy in containing the spread of problematic papers. (1) Changing the length of the tracking window produced qualitatively similar results (\si{Fig. S6}). (2) The finding was universally observed in each of four broad disciplines, including social sciences, life sciences, health sciences, and physical sciences (\si{Methods}; \si{Figs. S7-S10}).
(3) We obtained consistent results when excluding papers published in 9 multidisciplinary journals (\si{Figs. S13-S14}).
(4) Considering only uncritical tweets before retraction on Twitter shows once again the ineffectiveness of retraction in reducing attention (\si{Fig. S11}).
(5) We performed an interrupted time series analysis \cite{mcdowall2019interrupted} using the average weekly mention trajectory on all types of platforms and uncritical Twitter mentions to assess the effect of retraction on attention. This analysis confirms statistically that since online attention to retracted papers is exhausted when the retraction occurs, the retraction itself does not lead to a faster rate of decrease in mentions compared to the pre-retraction trend (\si{Table S6}).
(6) We investigated the fraction of mentions that occurred within the last month before retraction. We found that this fraction was very small for papers retracted more than seven months after their publication, which holds for both the overall attention (Figure~\ref{fig-last-month-frac}A) and the uncritical mentions on Twitter (Figure~\ref{fig-last-month-frac}B).  
These tests further support the main result that retractions do very little to limit the spread of problematic papers online, as attention has already been exhausted by the time retractions occur. This finding implies that retractions cannot be expected to remedy the problem that retracted papers get out-sized attention.

\section*{Discussion}

Our study shows that retracted papers attract more attention after publication than comparable non-retracted papers across a variety of online platforms including social and news media, blogs, and knowledge repositories. Moreover, their popularity surplus relative to non-retracted papers tends to be higher on curated than non-curated platforms. On the platform accountable for most mentions of research papers, retracted papers remain mentioned more often even after excluding critical tweets. These novel findings suggest that retracted papers are disseminated widely and through multiple channels before they are eventually retracted, possibly spreading flawed findings throughout the scientific community and the lay public.

Retracted papers might receive more attention due to a number of mechanisms.
First, problematic papers may naturally attract more attention.
For example, if the paper is retracted due to overclaiming, the results are likely to be presented as more significant, exciting, and attention-grabbing than they should.
Similarly, if the authors unintentionally make mistakes, they are more likely to publish the paper and receive attention if the results are positive rather than negative \cite{olson2002publication,easterbrook1991publication}. Hence, retracted papers are prone to present (false) positive findings and are therefore more eye-catching.
Second, attention leads to scrutiny which could increase the likelihood of retraction. 
In contrast, a paper that attracts limited attention may present little opportunity for retraction.

Our analysis of paper's critical mentions indicates that subsequently retracted papers are questioned relatively more often on Twitter after their publication than control papers.
This is nontrivial since retracted papers passed the peer-review system that science relies on to legitimize new findings through expert evaluation. 
Our finding validates at population-wide scale a recent case study of three retracted COVID-19/SARS-CoV-2 papers \cite{haunschild2021can} in suggesting that collective attention on Twitter includes meaningful discussions about science and may contain useful early signals related to problematic papers that could eventually contribute to their investigation.

We find that retracted papers continue to be discussed more than their control counterparts after being retracted. However, those discussions are primarily related to the retraction incident.
Surprisingly, retracted papers do not receive more mentions than control papers right before the retraction and the corresponding reference date.
This result is essential because it addresses an important open question about whether simply retracting a paper has an impact on the organic online dissemination of its findings, which can be extensive especially in the context of varied and broadly adopted online platforms as well as the increased engagement with science on most platforms \cite{scheufele2019science,hargittai2018young}. 
For the first time, we showed at scale that retraction has limited impact in reducing the spread of problematic papers online, as it comes after papers have already exhausted attention that is unaware of the retraction. This finding thus adds to knowledge about the consequences of retraction beyond the narrow scientific literature and academic sphere.
One practical implication is that journals, the scientific community, and the lay public should not think of retraction as an effective tool in decreasing online attention to problematic papers.

Although we studied the largest set of retracted papers to date, our work is not without limitations. First, not all problematic papers that should be retracted are retracted \cite{ioannidis2005most, fanelli2018opinion} and not all retracted papers are mentioned and traceable on major online platforms. Our findings are based solely on the subset recorded in the Altmetric database, which may not be representative of all problematic papers. 
Second, the Altmetric database may not cover all mentions of each paper. For instance, when a news outlet, such as the \emph{New Scientist}, posts a tweet with a link to its piece that covers a research paper, this tweet and its retweets would not necessarily be tracked by Altmetric unless they explicitly contain the unique identifier of the paper.
Third, our matching scheme controlling for journal topic, publication year, co-author team size, and author citation impact leaves other potential factors, such as prestige of affiliation, uncharted. Future work should thus expand on the matching used here and explore controlling for more factors that could impact attention volume.
Fourth, our detection of critical mentions is focused on a single platform. While Twitter accounts for a large fraction of all posts in the Altmetric database, future work needs to test our content-related findings on other online platforms.
Fifth, our study analyzes public attention to retracted papers. Distinguishing different types of online audiences including academics and professionals who engage with problematic papers is an interesting avenue for future research. 

Our results inform efforts concerned with the trustworthiness of science \cite{jamieson2019signaling,grimes2018modelling,fanelli2013us} and have crucial implications for the handling of retracted papers in academic publishing \cite{resnik2015retraction}. 
First, it is important to retract problematic papers as they are widely shared online by scholars, practitioners, journalists, and the public, some of whom might have no knowledge of their errors. 
Second, the fact that people tend to question these papers before retraction could be incorporated into efforts to identify questionable results earlier.
Third, our results show that by the time papers are retracted, they no longer receive much attention that is unaware of the retraction.
This suggests that the scientific community and the lay public should not expect journal retractions to be an effective tool to decrease harmful attention to problematic papers. 
Beyond revealing the limited efficacy of retractions in reducing attention to problematic papers, our work introduces a framework and contributes a public dataset of critical and uncritical mentions of research papers, which open up new avenues to study the dissemination of retracted papers beyond strictly academic circles.

\section*{Data and Methods}
\label{methods}

\subsection{Altmetric database}

The dataset that records online attention to research papers is provided by Altmetric (Oct 8, 2019 version). 
It monitors a range of online sources, searching their posts for links and references to published research.
The database we used contains posts for more than 26 million academic records which are primarily research papers.
Each post mentioning a paper contains a URL to its unique identifiers, such as Digital Object Identifier (DOI), PubMed ID, and arXiv ID.
Altmetric combines different identifiers for each unique paper, and also collates record of attention for different versions of each paper.
The database contains a variety of paper metadata, such as the title, authors, journal, publication date, and article type (research papers, books, book chapters, datasets, editorial pieces, etc.).
Our analysis was focused on research papers and their posts written in English.
For our critical-mention analysis, we collected all tweets referenced in the Altmetric database using the Twitter API. 
Due to users' privacy settings and account deletion, we successfully retrieved 90.7\% of the 85,411,606 tweets.
Detailed methods on defining four types of platforms, selecting top news outlets, and four broad disciplines are shown in \si{Methods}.

\subsection{Retraction Watch database}

We obtained the set of retracted articles from Retraction Watch \cite{rw}.
This database consists of 21,850 papers retracted by 2020.
Each paper has metadata including the title, journal, authors, publication and retraction date.
Out of the 9,201 retracted papers published after June 10, 2011, the launch date of \textit{Altmetric.com}, 8,434 also have DOI information. 
The majority of papers with missing DOI are conference abstracts.
We located 4,210 retracted papers in the Altmetric database based on DOI. 
Due to inconsistencies in recording and tracking retractions, we had to correct some publication and retraction dates. See details in \si{Methods}.

\subsection{Identifying control papers}


To compare the amount of attention received by retracted and non-retracted papers, we used the Altmetric database to construct a set of 5 control papers that had similar characteristics as retracted papers in terms of the publication year, publication venue, number of authors, and authors' total citation count one year before publication. 
The publication year controls for potential temporal variations in the amount of public attention to scientific papers online. 
Matching based on journal ensures that we are comparing papers from similar research areas, at least in the case of disciplinary venues, which might also affect the volume of attention. 
The number of authors and their citation status control for confounding factors related to prestige that can influence a paper's attention.
We collected information on author citations from the Microsoft Academic Graph \cite{sinha2015overview} based on papers' unique identifiers (DOI).
To ensure that we identified matching control papers for as many retracted papers as possible, we used coarsened exact matching~\cite{iacus2012causal}. 
For each retracted paper, we identified controls (i) published in the same journal, (ii) within 2 years of the publication year of the retracted paper (2 years before or 2 years after), (iii) that have no less or no more than 2 authors of the retracted paper, and (iv) whose most highly-cited author’s citation rank percentile is within 20\% of that of the retracted paper. Rank percentile was calculated based on all papers in the Altmetric database.
This matching strategy and its thresholds were determined based on a number of iterations that aimed to include as many potential confounding factors as possible, while finding matches for as many retracted papers as possible.
Retracted papers for which we found less than 5 matches were excluded. In total, we matched 19,255 control papers to 3,851 retracted papers.

\subsection{Defining tracking windows}

To ensure that we are using comparable time scales throughout our analyses, we defined tracking windows for the periods after publication, as well as before and after retraction. For the after publication time, we define the tracking window of retracted papers to be the time from their publication to their retraction date or the Altmetric database access date of October 9, 2019 (whichever comes first). 
We consider the tracking window of control papers to be the window from their publication date to the Altmetric database access date.
For the after retraction time period, we define a reference date for each paper. Retracted papers have their retraction date as their reference date. Control papers have the retraction date of their matched paper as their reference date.
The tracking window for both retracted and control papers is from the reference date to the Altmetric database access date.

\subsection{Labeling critical tweets}

We collected all pre-retraction tweets that mention retracted or control papers (control papers used the retraction date of their matched papers as the reference date). 
When attempting to identify critical tweets in this sample, we found that existing automated methods of detecting critical posts did not work well. 
We devised a novel approach that started with constructing a set of keywords to retrieve candidate tweets that may contain criticism. Our goal was for this heuristic to have high recall even if the precision was low. That is, we wanted the list to return a very large fraction of the critical tweets, even if it included many false positives. To construct the keyword list, we began with a set of seed words such as ``wonder'', ``concern'', ``suspect''. We then collected a random sample of 500 tweets and manually labeled them. Using these labels, we measured the recall of a simple criticism attribution based on whether the tweet contained any of the words from the current keyword list. As long as the recall was not high enough, we expanded our keyword list based on the content of tweets that were labeled as being critical but did not include any of the keywords. After two iterations, we achieved a recall of 0.9 with a list of 76 keywords (see \si{Methods}).
We applied this heuristic to 42,789 pre-retraction tweets and found 7,036 tweets that contained at least one of the keywords.

Then, we manually labeled each of these candidate tweets as critical or not. 
We used 3-4 annotators for this task. Each tweet was independently labeled by three different annotators. One of the authors trained the annotators by providing them a definition of critical tweets, examples, and labeling guidelines. The annotators were also familiarized with various aspects of scientific papers and publishing that are commonly criticized and questioned such as the review process, the validity of the data, the generalizability of the findings, etc. The annotators then labeled a batch of 500 tweets and discussed disagreements with one of the authors. They then labeled a second batch of 500 tweets, which resulted in an average Cohen's kappa score of 0.77, indicating substantial inter-rater agreement \cite{mchugh2012interrater}. The annotators then labeled all remaining tweets, resulting in 3 labels per tweet, achieving a weighted average Cohen's kappa score of 0.72.
Out of the 7,036 tweets, 6,201 had unanimous agreement among the 3 annotators, with 720 being labeled as critical. 
The other authors of the paper provided then a fourth label for the 835 tweets that had disagreement among the three annotators. After this step, 476 tweets had a majority label (3 vs. 1) and were labeled accordingly. The other 359 tweets remained ambiguous (2 vs. 2; \si{Table S7} shows some examples). We treated these ambiguous tweets as being uncritical. The results are qualitatively similar when treating them as critical (\si{Fig. S12}).


\subsection{Filtering retraction-related posts}

To label posts that discuss the retraction of a paper, we first identified all paper mentions that contained the term ``retract'', as it is often used in phrases such as ``it has been retracted'', ``journal retracts paper'', ``retraction of that paper'', etc. All such posts were considered to be discussing the retraction incident (a manual inspection of 50 random such posts on each type of platform shows no false positive cases). We then manually labeled each after-retraction post that did not contain this keyword as either discussing the retraction or not. 
Some examples are: ``We have to dig deeper! Science is never settled'' and ``Be careful when you are expected to believe something because `its science'.''
Using this method, we labeled all posts in news media and blogs. For social media, we labeled only posts from Twitter as this is the largest platform in that group. 
For knowledge repositories, we labeled Wikipedia posts, the second largest platform in this category (the largest one are patents, which we speculate do not usually discuss retraction). 
Note that since we did not label posts from all platforms, the mention volume of retracted papers after excluding retraction-related content is an overestimation.

\subsection{Data availability}

The Altmetric data can be accessed free of charge by researchers from \url{https://www.altmetric.com/research-access/}.
The Retraction Watch database is available freely from The Center For Scientific Integrity, subject to a standard data use agreement (see details at \url{https://retractionwatch.com/retraction-watch-database-user-guide/}.
The Microsoft Academic Graph is publicly available at \url{https://www.microsoft.com/en-us/research/project/microsoft-academic-graph/} or \url{https://www.microsoft.com/en-us/research/project/open-academic-graph/}.
We have uploaded all custom code and data created by us to the public Github repository at \url{https://github.com/haoopeng/retraction_attention/}.
The data we provide in our repository include unique identifiers for all retracted and control papers, which were used to link these data sources.
For user privacy reasons, Twitter does not allow sharing the text of tweets. To comply with this requirement, we provide IDs for the tweets that we labeled as critical/uncritical. Researchers can use these IDs to collect the text and metadata using the Twitter API (\url{https://developer.twitter.com/en/products/twitter-api/academic-research}).

\bibliography{references}

\begin{addendum} 
 \item[Acknowledgements] We thank \textit{Altmetric.com}, Retraction Watch, and Microsoft Academic Graph for providing the data used in this study. 
The authors thank Annika Weinberg, Berit Ginsberg, Henry Dambanemuya, and Rod Abhari for labelling critical tweets, Danaja Maldeniya for collecting the tweets used in this study, Jordan Braun for copy-editing, and Aparna Ananthasubramaniam, Danaja Maldeniya, and Ed Platt for helpful suggestions and discussion. This work has been partially funded by NSF CAREER Grant No IIS-1943506 and by the Air Force Office of Scientific Research under award number FA9550-19-1-0029.
 \item[Author Contributions] H.P., D.M.R., and E.\'A.H. designed the study; H.P. performed the analyses; H.P., D.M.R., and E.\'A.H. wrote the manuscript.
 \item[Competing Interests] The authors declare no competing interests.
 \item[Additional Information] Supplemental material is available for this paper. 
 \item[Materials \& Correspondence] Correspondence and requests for materials should be addressed to D.M.R. and E.\'A.H.
\end{addendum} 

\clearpage

\end{document}


\title{Dynamics of Cross-Platform Attention to Retracted Papers \\ (Supplemental Materials)}

\author{Hao Peng, Daniel M. Romero, Em\H oke-\'Agnes Horv\'at}

\date{\today}

\maketitle

\section{SI Methods}

\textbf{Four types of platforms.}
We categorized the 14 major platforms tracked by Altmetric into 4 groups: (1) social media (``twitter'', ``facebook'', ``googleplus'', ``linkedin'', ``pinterest'', ``reddit'', ``video''), (2) news media (``news media''), (3) blogs (``blogs''), and (4) knowledge repositories (``wikipedia'', ``patent'', ``f1000'', ``q\&a'', ``peer reviews''). Three platforms (``policy'', ``misc'', and ``weibo'') were excluded as they do not fit into these categories and make up only a tiny fraction of mentions (less than 1\%).
During data cleaning, we relabelled as ``news media'' 30 mislabelled ``blogs'' (out of 1,262), such as ``Scientific American'', ``Medical Daily'', and ``Popular Science''.\\

\textbf{Top news outlets.}
We compiled a list of 27 top news outlets that are believed to be trustworthy, have high-quality science reporting, and represent popular media brands across a range of platforms according to the Pew Research Center \cite{taneja2019people,jurkowitz2020us}. These outlets (hereafter referred to as ``top news'') are: \emph{ABC News}, \emph{BBC News}, \emph{Breitbart News Network}, \emph{Business Insider}, \emph{Buzzfeed}, \emph{CBS News}, \emph{CNN News}, \emph{FOX News}, \emph{Huffington Post}, \emph{MSNBC}, \emph{NBC News}, \emph{NPR}, \emph{New York Post}, \emph{New York Times}, \emph{Newsweek}, \emph{PBS}, \emph{Politico Magazine}, \emph{TIME Magazine}, \emph{The Daily Caller}, \emph{The Guardian}, \emph{The Hill}, \emph{USA Today}, \emph{Vice}, \emph{Vox.com}, \emph{Wall Street Journal}, \emph{Washington Examiner}, and \emph{Washington Post}.\\

\textbf{Four broad disciplines.}
The Altmetric database also provides the research fields for each paper based on the 26 Scopus Subject Areas, which belong to 4 broad disciplines including Social Sciences, Life Sciences, Physical Sciences, and Health Sciences \cite{scopus}. The subject classification was performed by in-house experts based on the aim and scope of the content a journal publishes. Note that a paper can belong to several subject areas and therefore multiple disciplines.\\

\textbf{Establishing publication date.}
Our analysis relies on identifying the correct publication and retraction date for each paper.
A paper can have different online and print publication dates (online is usually earlier). 
The Altmetric database provides both ``pubdate'' and ``epubdate'' for each paper. The Retraction Watch database also provides the publication date of retracted papers.
We thus set the publication date for each retracted paper as the earliest date recorded by the two databases.\\

\textbf{Correcting retraction date.}
Similarly, the retraction date can be considered to be the online or the print version of the retraction notice. For example, the official retraction announcement for the paper ``Molecular Characterization and Biological Activity of Interferon-$\alpha$ in Indian Peafowl (Pavo cristatus)'' \cite{zhao2017retracted} was published in print on October 1, 2017, while it was posted online already on September 22, 2017.
The retraction date recorded in Retraction Watch contains some inconsistencies, which could influence our analysis of pre- and post-retraction attention if not resolved.
To ensure that we accurately separate mentions posted before the retraction from those posted after it, we utilize Twitter posts, the largest single platform in terms of attention volume (nearly 80\%), to correct the retraction date for each paper. Our assumption is that if any tweet mentions the retraction of a paper (e.g., ``.@PLOSONE announced \#retraction of this @XYZ paper, notice will come in 2 weeks. [URL]'') posted before the recorded retraction date, the effective retraction date should be at least as early as the time of this tweet.

We first identified 572 tweets posted before each paper's retraction date (as per the Retraction Watch database) such that the tweets contain the string ``retract'' in its lowercase text. 
Not all such tweets are announcing a retraction (e.g., ``Please retract this paper!!!'' and ``How is it possible that this study hasn't been retracted?'').
Hence, we manually labeled these tweets and found that 469 of them contained a statement of retraction. 
We corrected the retraction date for 381 papers based on this set of labeled tweets. A manual inspection of a random sample of 20 papers from the remaining retracted papers shows no error in the retraction date recorded in the Retraction Watch database.\\


\textbf{76 criticism keywords}
Our iterative process to develop a heuristic for pre-selecting potentially critical tweets has resulted in the following keywords:\\

\textit{verify, problem, really, pretend, fake, raise, believe, story, possible, data, concern, crackpot, intelligence, standard, correction, possibly, wonder, laughable, stupid, distrust, pseudoscience, conclusive, insult, surprise, dumb, unclear, peer review, suspect, criticism, suspicious, replication, unsure, manipulation, ?, validate, plausible, controversy, shame, plausibly, non-expert, confounding, confusing, serious, academic, joke, chance, question, shock, fraud, controversial, nonsense, uncertain, manipulated, replicate, embarrass, verified, confounder, validation, peer-review, claim, publish, not sure, lol, intelligent, reputable, reviewer, reputation, warrant, lying, doubt, lie, academy, flaw, retract, believable, confuse.}

\section{SI Figures}


\begin{figure*}[ht!] 
\centering
\includegraphics[trim=0mm 0mm 0mm 0mm, width=0.9\linewidth]{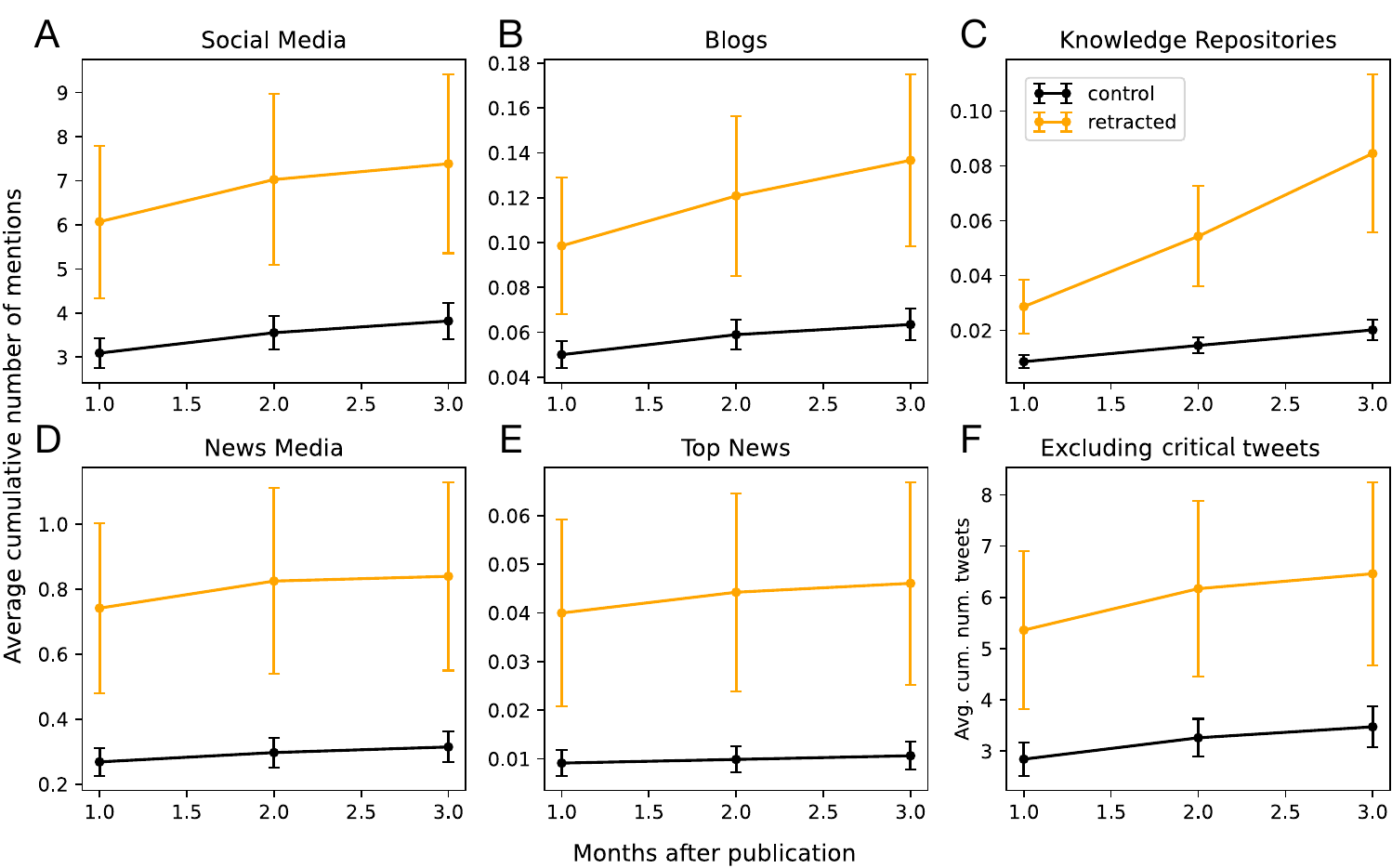}
\caption{\textbf{After-publication mentions.} \textbf{A-E}, Average cumulative number of mentions received within 3 months after publication on four types of platforms and in top news outlets for both retracted and control papers. \textbf{F}, Average cumulative number of mentions on Twitter after excluding critical tweets. Comparisons across different types of platforms show that retracted papers receive more attention after publication than non-retracted papers. Error bars indicate 95\% confidence intervals.}
\label{fig-after-pub-3m}
\end{figure*} 

\begin{figure*}[ht!] 
\centering
\includegraphics[trim=0mm 0mm 0mm 0mm, width=0.4\linewidth]{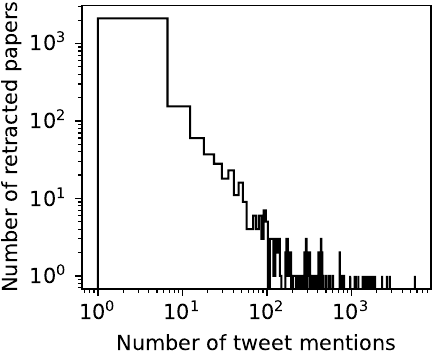}
\caption{\textbf{Distribution of the number of mentions on Twitter for retracted papers.} The distribution of online attention to retracted papers on Twitter is right-skewed as only a few papers can become very popular.}
\label{fig-ret-tw-dist}
\end{figure*} 

\begin{figure*}[ht!] 
\centering
\includegraphics[trim=0mm 0mm 0mm 0mm, width=0.5\linewidth]{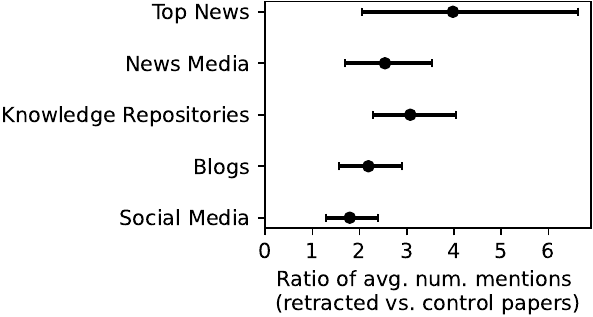}
\caption{Ratio of the average number of mentions of retracted papers to that of non-retracted papers on each type of platform. Relative to non-retracted papers, retracted ones tend to be favored more in the amount of coverage on news media, including top news, and knowledge repositories than on social media and blogs. However, the differences are not statistically significant due to imprecise estimation. Error bars indicate 95\% bootstrapped confidence intervals.}
\label{fig-platform-ratio}
\end{figure*} 

\begin{figure*}[ht!] 
\centering
\includegraphics[trim=0mm 0mm 0mm 0mm, width=0.6\linewidth]{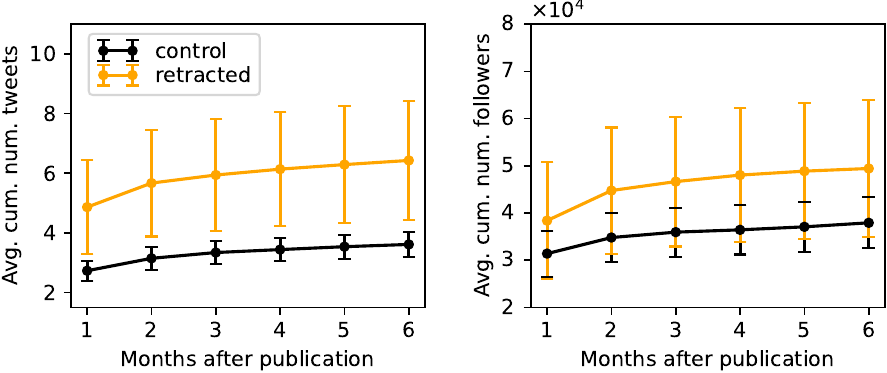}
\caption{\textbf{Number of followers for tweets that mention research papers within a given number of months after their publication.} \textit{Left}, Average cumulative number of tweets received within 6 months after publication by retracted and control papers. \textit{Right}, Same as \textit{Left} but extended to the number of followers of those tweets.}
\label{fig-after-pub-fw}
\end{figure*} 

\begin{figure*}[ht!] 
\centering
\includegraphics[trim=0mm 0mm 0mm 0mm, width=\linewidth]{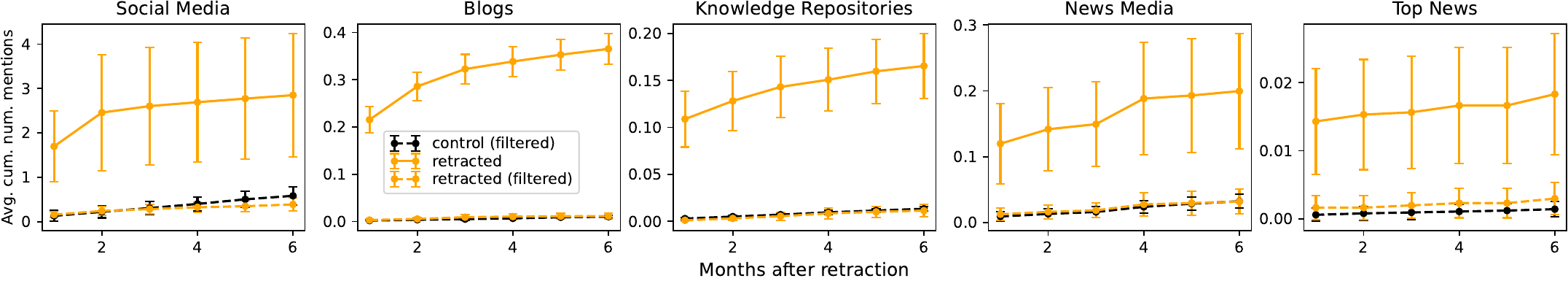}
\caption{\textbf{After-retraction mentions.} \textbf{A-E}, Average number of mentions received within 6 months after retraction on four types of platforms and in top news outlets for retracted and control papers. 
Comparisons across different types of platforms show that retracted papers receive more attention even after their retraction than non-retracted papers (solid orange lines). However, excluding retraction-related posts eliminates the surplus of attention that retracted papers receive (dashed orange lines).
Error bars indicate 95\% confidence intervals.}
\label{fig-after-ret-si-6m}
\end{figure*} 

\begin{figure*}[ht!] 
\centering
\includegraphics[trim=0mm 0mm 0mm 0mm, width=\linewidth]{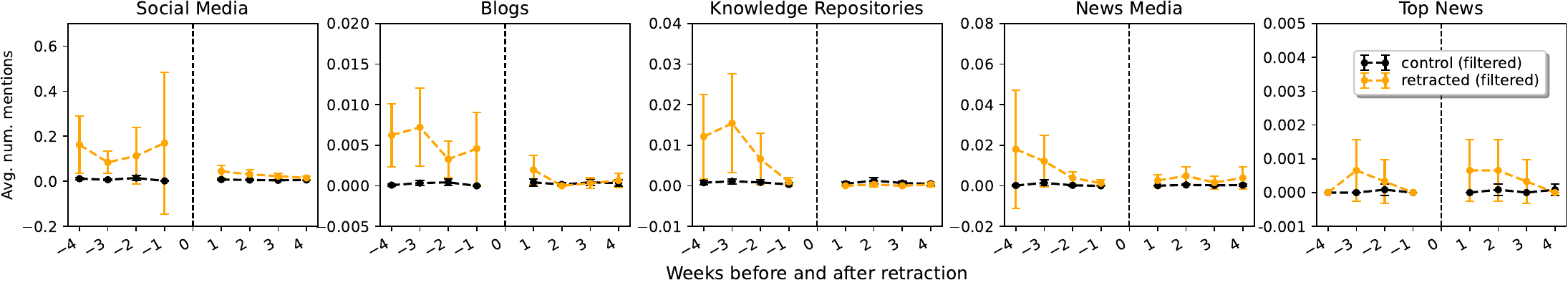}
\caption{\textbf{Average number of weekly mentions within 1 month before and after retraction.} Trends are shown on four types of platforms and in top news outlets for both retracted and control papers. 
In the case of retracted papers, we excluded all after-retraction posts that discussed the retraction in some form. For consistency, we also excluded posts that contain the phrase ``retract'' when referring to both control and retracted papers in the whole time period. 
Comparisons across different types of platforms show that retracted papers do not receive statistically more mentions than control papers immediately before or after retraction.
Error bars indicate 95\% confidence intervals.}
\label{fig-before-after-ret-1m}
\end{figure*} 

\begin{figure*}[ht!] 
\centering
\includegraphics[trim=0mm 0mm 0mm 0mm, width=\linewidth]{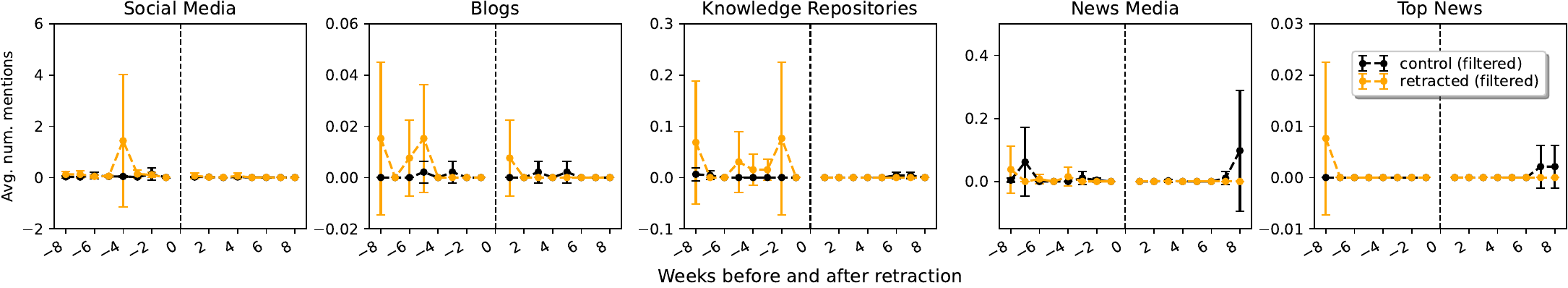}
\caption{\textbf{Average number of weekly mentions within 2 months before and after retraction.} Trends are shown on four types of platforms and in top news outlets for 131 retracted and 470 control papers in \textbf{Social Sciences}.
In the case of retracted papers, we excluded all after-retraction posts that discussed the retraction in some form. For consistency, we also excluded posts that contain the phrase ``retract'' when referring to both control and retracted papers in the whole time period. 
Comparisons across different types of platforms show that retracted papers do not receive statistically more mentions than control papers immediately (e.g., within two weeks) before or after retraction.
Error bars indicate 95\% confidence intervals.}
\label{fig-before-after-ret-2m-social}
\end{figure*} 

\begin{figure*}[ht!] 
\centering
\includegraphics[trim=0mm 0mm 0mm 0mm, width=\linewidth]{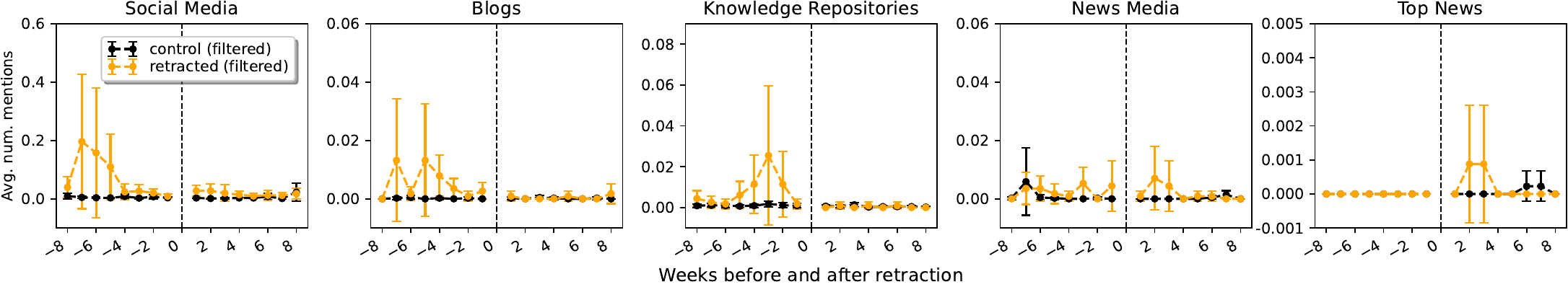}
\caption{Same as Fig.~\ref{fig-before-after-ret-2m-social}, but for 1,137 retracted and 4,423 control papers in \textbf{Life Sciences}.}
\label{fig-before-after-ret-2m-life}
\end{figure*} 

\begin{figure*}[ht!] 
\centering
\includegraphics[trim=0mm 0mm 0mm 0mm, width=\linewidth]{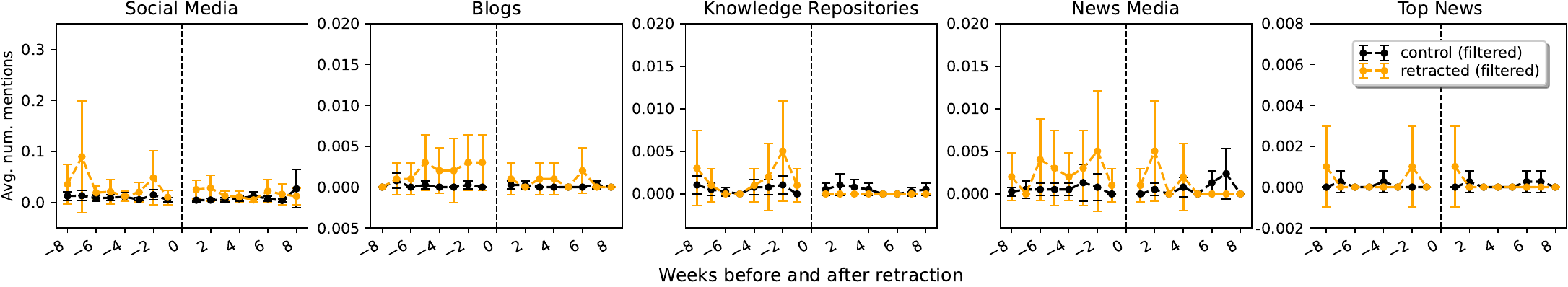}
\caption{Same as Fig.~\ref{fig-before-after-ret-2m-social}, but for 993 retracted and 3,780 control papers in \textbf{Health Sciences}.}
\label{fig-before-after-ret-2m-health}
\end{figure*} 

\begin{figure*}[ht!] 
\centering
\includegraphics[trim=0mm 0mm 0mm 0mm, width=\linewidth]{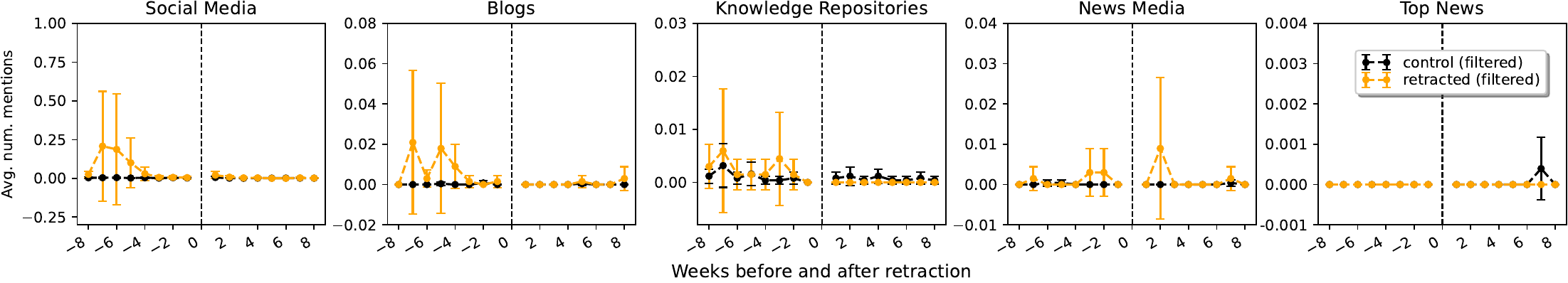}
\caption{Same as Fig.~\ref{fig-before-after-ret-2m-social}, but for 669 retracted and 2,524 control papers in \textbf{Physical Sciences}.}
\label{fig-before-after-ret-2m-physical}
\end{figure*} 

\begin{figure*}[ht!] 
\centering
\includegraphics[trim=0mm 0mm 0mm 0mm, width=0.4\linewidth]{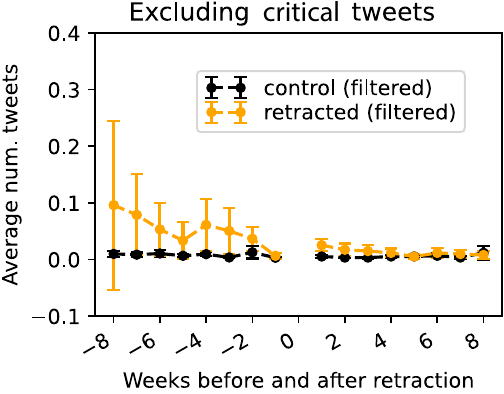}
\caption{\textbf{Before- and after-retraction mentions on Twitter.} Average weekly number of tweets received within 2 months before and after retraction for both retracted and control papers. 
We excluded after-retraction tweets that discuss the retraction itself for retracted papers. 
We also excluded tweets that contain the phrase ``retract'' for both groups of papers in the whole time period.
For all papers, we removed critical tweets before retraction.
Error bars indicate 95\% confidence intervals.
This result shows that retracted papers are discussed not statistically more often than control papers right before their retraction.
}
\label{fig-before-after-ret-tw}
\end{figure*} 

\begin{figure*}[ht!] 
\centering
\includegraphics[trim=0mm 0mm 0mm 0mm, width=0.35\linewidth]{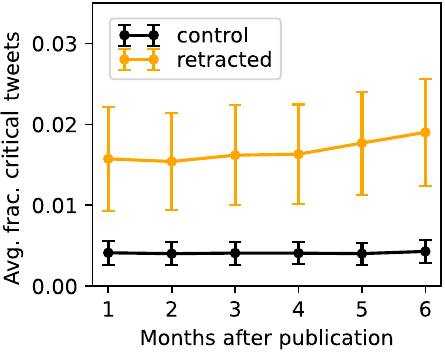}
\caption{\textbf{Average cumulative fraction of critical tweets within 6 months after publication.} 
We treat ambiguous tweets as expressing critical concerns.
Similar to Fig. 3 shown in the main paper, retracted papers receive a higher fraction of critical tweets. 
Error bars indicate 95\% confidence intervals.}
\label{fig-after-pub-uncrtn-bar}
\end{figure*} 

\begin{figure*}[ht!] 
\centering
\includegraphics[trim=0mm 0mm 0mm 0mm, width=0.9\linewidth]{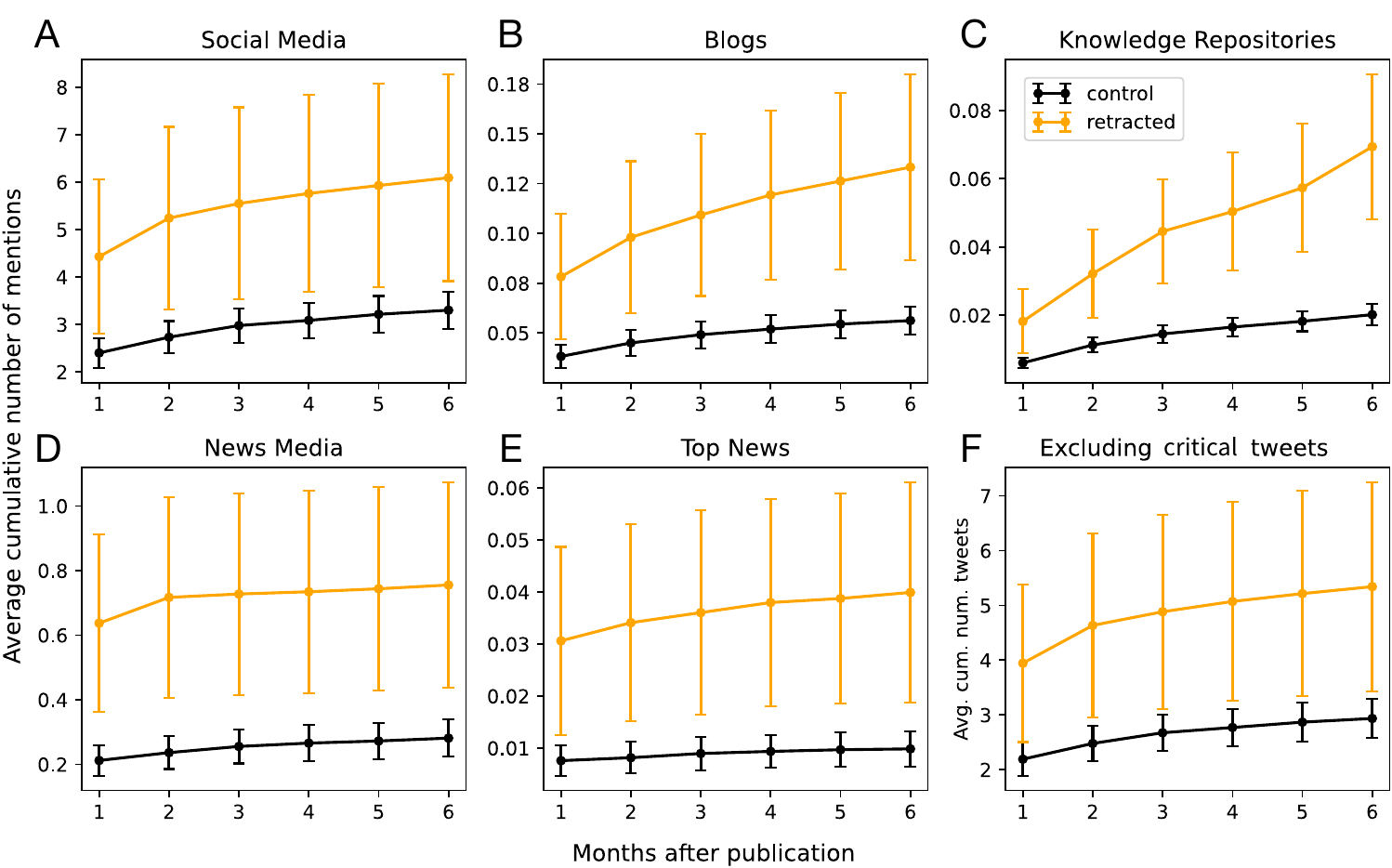}
\caption{Same as Fig. 2 in the main text, but after excluding 281 retracted papers (and their control papers) published in 9 multidisciplinary journals, including Nature, Science, PNAS, Nature Communications, Science Advances, Scientific Reports, PLoS One, Philosophical Transactions of the Royal Society A, and Proceedings of the Royal Society A. We compiled this list of general science journals as they are prominent journals classified as multidisciplinary in most journal classification systems \cite{borner2012design}. Error bars indicate 95\% confidence intervals.}
\label{fig-after-pub-6m-exc-general-j}
\end{figure*} 

\begin{figure*}[ht!] 
\centering
\includegraphics[trim=0mm 0mm 0mm 0mm, width=\linewidth]{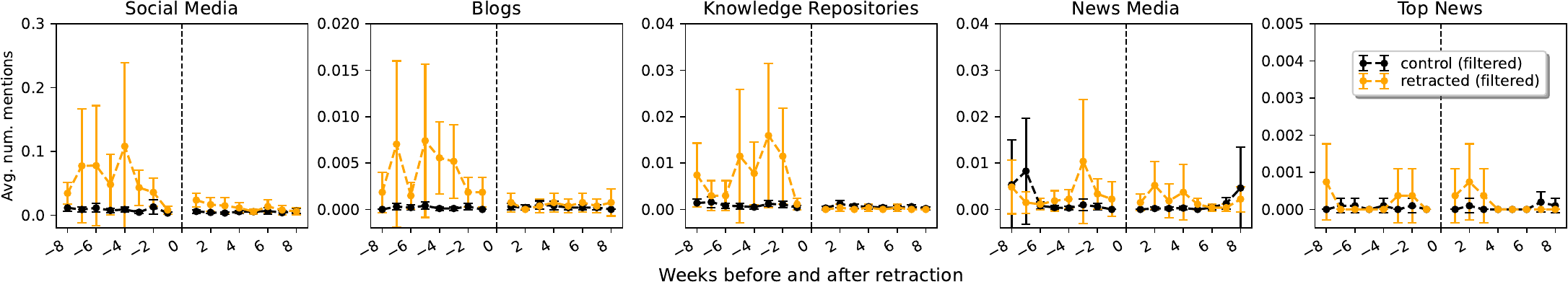}
\caption{
Same as Fig. 4 in the main text, but after excluding 281 retracted papers and all their control papers published in 9 multidisciplinary journals. Error bars indicate 95\% confidence intervals.}
\label{fig-before-after-ret-2m-exc-general-j}
\end{figure*} 

\begin{figure*}[h!] 
\centering
\includegraphics[trim=0mm 0mm 0mm 0mm, width=0.35\columnwidth]{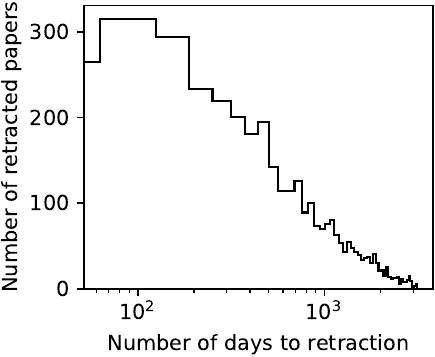}
\caption{The histogram of the number of days to retraction for all retracted papers used in our analysis. The median time to retraction is 480 days, and the average time to retraction is 698 days.}
\label{fig:gap}
\end{figure*} 

\begin{figure*}[ht!] 
\centering
\includegraphics[trim=0mm 0mm 0mm 0mm, width=0.4\linewidth]{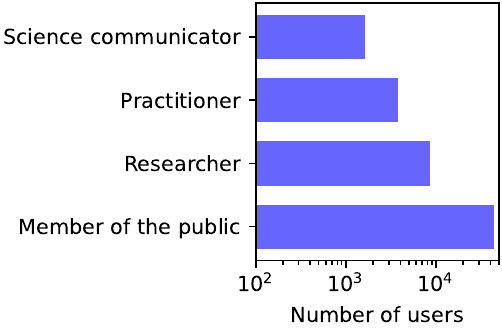}
\caption{Frequency of four types of users on Twitter who have mentioned retracted papers. The user classification is provided by in-house experts at Altmetric.}
\label{audience-type}
\end{figure*} 

\clearpage

\section{SI Tables}

\begin{table*}[ht]
\centering
\begin{tabular}{|l|l|l|l|l|l|}
\hline
 \multicolumn{1}{|c|}{Platform type} & \multicolumn{1}{c|}{Time window} & \multicolumn{1}{c|}{Retraction coeff.} & \multicolumn{1}{c|}{Pub. year coeff.} & \multicolumn{1}{c|}{Num. authors coeff.} & \multicolumn{1}{c|}{Log citation coeff.}\\ \hline
Social Media  &  1  &  0.583***  &  0.310***  &  0.039***  &  0.178***  \\ \hline
Social Media  &  2  &  0.607***  &  0.309***  &  0.036***  &  0.178***  \\ \hline
Social Media  &  3  &  0.583***  &  0.300***  &  0.032***  &  0.180***  \\ \hline
Social Media  &  4  &  0.581***  &  0.297***  &  0.031***  &  0.177***  \\ \hline
Social Media  &  5  &  0.572***  &  0.296***  &  0.032***  &  0.175***  \\ \hline
Social Media  &  6  &  0.568***  &  0.295***  &  0.032***  &  0.171***  \\ \hline
Blogs  &  1  &  0.735***  &  0.125***  &  0.050***  &  0.306***  \\ \hline
Blogs  &  2  &  0.776***  &  0.121***  &  0.054***  &  0.287***  \\ \hline
Blogs  &  3  &  0.803***  &  0.115***  &  0.052***  &  0.279***  \\ \hline
Blogs  &  4  &  0.835***  &  0.120***  &  0.051***  &  0.269***  \\ \hline
Blogs  &  5  &  0.840***  &  0.113***  &  0.051***  &  0.259***  \\ \hline
Blogs  &  6  &  0.856***  &  0.110***  &  0.051***  &  0.253***  \\ \hline
Knowledge Repositories  &  1  &  1.175***  &  -0.071  &  0.086***  &  0.215***  \\ \hline
Knowledge Repositories  &  2  &  1.079***  &  -0.105***  &  0.081***  &  0.189***  \\ \hline
Knowledge Repositories  &  3  &  1.066***  &  -0.118***  &  0.089***  &  0.169***  \\ \hline
Knowledge Repositories  &  4  &  1.092***  &  -0.113***  &  0.095***  &  0.161***  \\ \hline
Knowledge Repositories  &  5  &  1.127***  &  -0.117***  &  0.090***  &  0.159***  \\ \hline
Knowledge Repositories  &  6  &  1.166***  &  -0.099***  &  0.085***  &  0.161***  \\ \hline
News Media  &  1  &  1.002***  &  0.388***  &  0.050***  &  0.372***  \\ \hline
News Media  &  2  &  0.988***  &  0.387***  &  0.050***  &  0.354***  \\ \hline
News Media  &  3  &  0.930***  &  0.389***  &  0.048***  &  0.346***  \\ \hline
News Media  &  4  &  0.904***  &  0.389***  &  0.046***  &  0.348***  \\ \hline
News Media  &  5  &  0.896***  &  0.394***  &  0.045***  &  0.347***  \\ \hline
News Media  &  6  &  0.875***  &  0.395***  &  0.046***  &  0.339***  \\ \hline
Top News  &  1  &  1.469***  &  0.283***  &  0.027*  &  0.416***  \\ \hline
Top News  &  2  &  1.497***  &  0.292***  &  0.030*  &  0.397***  \\ \hline
Top News  &  3  &  1.458***  &  0.279***  &  0.029*  &  0.397***  \\ \hline
Top News  &  4  &  1.463***  &  0.283***  &  0.025*  &  0.399***  \\ \hline
Top News  &  5  &  1.452***  &  0.285***  &  0.022  &  0.400***  \\ \hline
Top News  &  6  &  1.464***  &  0.282***  &  0.024*  &  0.398***  \\ \hline
\end{tabular}
\caption{Association between retraction and the number of mentions within a given number of months after publication. Models control for the publication year, number of authors, the most highly-cited author's log citation count, and are set up to predict the number of mentions in each type of platform. Negative binomial regression coefficients indicate statistically significant correlations between retraction and the number of mentions. Each row corresponds to each combination of platform type and time window (in months).
The coefficients reveal that, compared to control papers, retracted papers receive more mentions on knowledge repositories and top news than on social media and blogs.
It also shows that the magnitude of association between retraction and mentions is relatively stable over time on social media, knowledge repositories, and in top news.
However, this association is gradually decreasing on news media, indicating that retracted papers enjoy a surplus of popularity mostly in the initial stage of dissemination, and that they experience a faster attention decay on news media than control papers. 
In contrast, the attention gap between two groups of papers is increasing on blogs. 
Differences in temporal patterns on the investigated platforms may be related to different spreading mechanisms and platform audiences, which deserves further investigation in future work.
Significance levels: *** p$<$0.001, ** p$<$0.01, and * p$<$0.05.}
\label{tab-after-pub}
\end{table*}

\begin{table*}[ht]
\centering
\begin{tabular}{|l|l|l|l|}
\hline
  \multicolumn{1}{|c|}{Platform type} & \multicolumn{1}{c|}{Mean value (retract vs. control)} & \multicolumn{1}{c|}{P-value/T-test} & \multicolumn{1}{c|}{P-value/U-test} \\ \hline
Social Media & 7.180 vs. 4.011 & 0.000 & 0.000 \\ \hline
Blogs & 0.160 vs. 0.073 & 0.000 & 0.004 \\ \hline 	 
Knowledge Repositories & 0.088 vs. 0.029 & 0.000 & 0.000 \\ \hline
News Media & 0.845 vs. 0.335 & 0.000 & 0.673 \\ \hline
Top News & 0.046 vs. 0.012 & 0.000 & 0.000 \\ \hline
\end{tabular}
\caption{The average total number of mentions within 6 months after publication on each type of platform. The T-test for the means of two groups of papers shows that retracted papers receive a statistically higher average number of mentions than control papers. 
The U-test shows that the central tendency of the distribution of mention counts for retracted papers is larger than that of control papers, except for News Media.}
\label{u-test}
\end{table*}

\begin{table*}[ht]
\centering
\begin{tabular}{|l|l|l|l|l|l|}
\hline
 \multicolumn{1}{|c|}{Platform type} & \multicolumn{1}{c|}{Time window} & \multicolumn{1}{c|}{Retraction coeff.} & \multicolumn{1}{c|}{Pub. year coeff.} & \multicolumn{1}{c|}{Num. authors coeff.} & \multicolumn{1}{c|}{Log citation coeff.}\\ \hline
Twitter  &  1  &  0.546***  &  0.333***  &  0.038***  &  0.185***  \\ \hline
Twitter  &  2  &  0.560***  &  0.333***  &  0.034***  &  0.185***  \\ \hline
Twitter  &  3  &  0.542***  &  0.328***  &  0.032***  &  0.185***  \\ \hline
Twitter  &  4  &  0.543***  &  0.324***  &  0.032***  &  0.181***  \\ \hline
Twitter  &  5  &  0.536***  &  0.324***  &  0.032***  &  0.178***  \\ \hline
Twitter  &  6  &  0.536***  &  0.322***  &  0.032***  &  0.176***  \\ \hline
\end{tabular}
\caption{Negative binomial regression coefficients for variables: retraction status, publication year, number of authors, and author's log citations. The dependent variable is the number of mentions on Twitter (excluding critical tweets) within a given time window of publication.
Significance levels: *** p$<$0.001, ** p$<$0.01, and * p$<$0.05.}
\label{tab-after-pub-tw-exclude-uncertain}
\end{table*}

\begin{table*}[ht]
\centering
\begin{tabular}{|l|l|l|l|l|l|}
\hline
  \multicolumn{1}{|c|}{Platform type} & \multicolumn{1}{c|}{Time window} & \multicolumn{1}{c|}{Retraction coeff.} & \multicolumn{1}{c|}{Pub. year coeff.} & \multicolumn{1}{c|}{Num. authors coeff.} & \multicolumn{1}{c|}{Log citation coeff.}\\ \hline
Twitter  &  1  &  0.002**  &  0.000  &  0.000  &  0.000  \\ \hline
Twitter  &  2  &  0.002**  &  0.000*  &  0.000  &  -0.000  \\ \hline
Twitter  &  3  &  0.002**  &  0.000  &  0.000  &  -0.000  \\ \hline
Twitter  &  4  &  0.002**  &  0.000  &  0.000  &  -0.000  \\ \hline
Twitter  &  5  &  0.003***  &  0.000*  &  0.000  &  0.000  \\ \hline
Twitter  &  6  &  0.004***  &  0.000**  &  0.000  &  -0.000  \\ \hline
\end{tabular}
\caption{Linear regression coefficients for the variables: retraction status, publication year, number of authors, and author's log citations. The dependent variable is the fraction of critical tweets received after publication within a given time window. Significance levels: *** p$<$0.001, ** p$<$0.01, and * p$<$0.05.}
\label{tab-after-pub-tw-frac-uncertain}
\end{table*}

\begin{table*}[ht]
\centering
\begin{tabular}{|l|l|l|}
\hline
  \multicolumn{1}{|c|}{Platform type} & \multicolumn{1}{c|}{Correlation coefficient} & \multicolumn{1}{c|}{P-value} \\ \hline
Social Media & 0.031 & 0.30 \\ \hline
Blogs & 0.019 & 0.52 \\ \hline 	 
Knowledge Repositories & 0.005 & 0.86 \\ \hline
News Media & -0.004 & 0.91 \\ \hline
Top News & 0.002 & 0.95 \\ \hline
\end{tabular}
\caption{Pearson correlation coefficient between the fraction of critical tweets and the total number of mentions on each type of platform separately, received within 6 months after publication. This analysis focuses on retracted papers that have at least one tweet mention.}
\label{tab-tw-frac-uncertain-correlation}
\end{table*}

\begin{table*}[]
\begin{tabular}{|c|c|c|c|c|c|c|}
\hline
\multicolumn{1}{|l|}{} & Social Media & Blogs   & Knowledge Repositories & News Media & Top News  & Twitter    \\ \hline
X1                     & -0.013***    & -0.0001 & 0.0005                 & -0.000008  & -0.000004 & -0.0097*** \\ \hline
X2                     & 0.0013       & -0.0032 & -0.0093                & -0.0015    & 0.0003    & 0.0044     \\ \hline
X3                     & 0.01*        & 0.00007 & -0.0004                & -0.0003    & -0.00008  & 0.0077**   \\ \hline
Intercept              & 0.1325***    & 0.005** & 0.0057                 & 0.0046*    & 0.0002    & 0.0959***  \\ \hline
\end{tabular}
\caption{Coefficients of the interrupted time series analysis \cite{mcdowall2019interrupted} using the average weekly number of mentions on each type of platform for retracted papers (2 months before and 2 months after retraction). 
For all types of platforms, we excluded retraction-related posts after retraction, as well as posts containing the phrase ``retract'' throughout. 
For Twitter, we additionally excluded critical tweets before retraction.
$X1$ represents the attention trend before retraction. It is negative and significant for Social Media (and Twitter), indicating that the attention is decreasing over time.
$X3$ represents the difference between before- and after retraction slopes. It is positive and significant for Social Media (and Twitter), indicating that the attention volume is still decreasing, but the magnitude has decreased.
$X2$ represents the change in attention immediately after retraction, which is not significant on any type of platform, suggesting that retraction has no immediate effect on attention change. Significance levels: *** p$<$0.001, ** p$<$0.01, and * p$<$0.05.}
\label{ITS-all-plat}
\end{table*}


\begin{table*}[ht]
\centering
\begin{tabular}{|r|}
\hline
Lol! What a conclusion. [url]                                                                                                                      \\ \hline
\begin{tabular}[c]{@{}r@{}}Darlings, another mediterranean diet study claiming lowfat, not \\ lowfat that proves nothing. fat=30\% [url] \#vegan \#dietitian\end{tabular} \\ \hline
\begin{tabular}[c]{@{}r@{}}I can't believe I haven't seen anyone in my \\ feed mentioning this grade. A bull shit [url]\end{tabular}                                             \\ \hline
\begin{tabular}[c]{@{}r@{}}Isn't osmium a tad toxic to be an antitumour agent? \\ (I haven't read the paper yet...) - [url]\end{tabular}                                    \\ \hline
\begin{tabular}[c]{@{}r@{}}Religion and altruism correlated. How exactly is \\ this current biology? [url]\end{tabular}                                                         \\ \hline
\end{tabular}
\caption{Five examples of ambiguous tweets in labelling critical tweets. Note that the content of each tweet has been paraphrased for anonymization.}
\label{uncrtn-tw-tab}
\end{table*}

\clearpage

\bibliographystyle{plainnat}
\bibliography{references}